\documentclass{aa}  

\usepackage{graphicx}
\usepackage{txfonts}
\usepackage{amsmath}	% Advanced maths commands
\usepackage{newtxtext,newtxmath}
\usepackage{graphicx}% Include figure files
\usepackage{dcolumn}% Align table columns on decimal point
\usepackage{bm}% bold math
\usepackage{mathtools}
\usepackage{float}
\usepackage{linenoaa}
\usepackage{caption}
\usepackage{booktabs}
\usepackage{array}
\usepackage{geometry}
\usepackage{subcaption}
\usepackage{physics}
\usepackage[dvipsnames]{xcolor}
\usepackage{natbib}
\usepackage{hyperref}
\usepackage{ulem}
\definecolor{argie}{RGB}{106, 13, 173}

\hypersetup{colorlinks = true,
	allcolors = blue,}
\bibpunct{(}{)}{;}{a}{}{,}
\begin{document} 
	
	\titlerunning{Detachment Instability in MADs}
	\authorrunning{A. Loules et al.}
    % Exploring the Possibility of a Detachment Instability in Magnetically Arrested Disks
	\title{The azimuthal structure of magnetically arrested disks during flux eruption events}
	%MADs\sout{The Parker instability in 
		\subtitle{}
		
		\author{Argyrios Loules\inst{1},
			Antonios Nathanail\inst{2},
			\and
			Ioannis Contopoulos\inst{2}
		}
		
		\institute{Section of Astrophysics, Astronomy and Mechanics, Department of           Physics, National and Kapodistrian University of Athens,
			University Campus, Zografos GR-15784, Athens, Greece\\
			\email{arloules@phys.uoa.gr}
			\and
			Research Center for Astronomy and Applied Mathematics, Academy of Athens, Soranou Efessiou 4, GR-11527 Athens, Greece\\
			\email{anathanail@academyofathens.gr, icontop@academyofathens.gr}
		}

		\date{}
		
		\abstract
    {Magnetically arrested disks (MADs) are highly dynamic astrophysical systems characterized by strong variability and transient phenomena such as magnetic flux eruption events.}
    {In this work, we investigate the azimuthal structure of the equatorial inner accretion flow during flux eruption events, and propose a physical mechanism for the formation and outward transport of vertical magnetic flux tubes.}
    {We analyze data from a standard 3D general-relativistic magnetohydrodynamics (GRMHD) simulation, focusing on equatorial slices in order to examine the details and the evolution of the azimuthal structure of the accreting matter.}
    {During flux eruption events, the non-axisymmetric features of the equatorial inner accretion disk are considerably enhanced, with this enhancement being more prominent close to the black hole. Our analysis of the azimuthal structure of the equatorial accretion disk finds that the matter distribution in the vicinity of the horizon is dominated by low azimuthal mode numbers, specifically by the $m = 2$, and $m = 1$ modes, indicating that the non-axisymmetry of the disk during flux eruption events is enhanced due to the emergence of features with a large angular size on the equatorial plane. Our results suggest that the morphology of the equatorial accretion flow close to the black hole is mainly determined by the formation and motion of vertical magnetic flux bundles. These bundles are formed when the initially horizontal magnetic field reconnects into a vertical configuration, effectively detaching from the black hole horizon. This reconnection occurs in a low-density, highly magnetized region on the equatorial plane that expands over time as more field lines undergo vertical reconfiguration. The resulting vertical flux tubes, filled with low-density plasma, are then transported outwards due to magnetic buoyancy.}
    {Our results present a detailed quantitative description of the morphology of MADs and of its evolution during flux eruptions, complemented by a description of the physical process by which excess magnetic flux is detached from the black hole, vertically reconfigured, and expelled.}
% 5 {} token are mandatory
		
		
		\maketitle
		\nolinenumbers
		%
		%-------------------------------------------------------------------
		
		\section{Introduction}
		
The accretion of matter onto black holes lies at the heart of most high-energy astrophysical environments, as the engine that powers intense cosmic processes, such as the launching of relativistic jets of magnetized plasma \citep{blandford1977, begelman1984}. Due to its central role in shaping the dynamical behavior of these most extreme of environments in Nature, a significant amount of literature, over the span of more than four decades, has been devoted to uncovering the details of matter accretion onto black holes. 

The accreting material is typically thought to assume the configuration of a thin disk \citep{novikov1973, shakura1973, pringle1973} or of a geometrically thick torus \citep{fishbone1976, abramowicz1978} around the black hole. As it accretes, its magnetic field, which is frozen into the plasma, is advected towards the black hole, with magnetic flux in the vicinity of the event horizon steadily increasing until the accumulated magnetic field is strong enough to counteract the ram pressure of the infalling gas. Accretion then proceeds intermittently, in the highly transient MAD state \citep{bisnovatyi-kogan1974, bisnovatyi-kogan1976, narayan2003}, which has been confirmed by numerical simulations \citep{igumenshchev2008}, and is in agreement with the ordered magnetic fields observed in M87* \citep{ehtc2021} and Sgr A* \citep{ehtc2024}.

The MAD state, however, appears not to be exclusive to the accretion of a Fishbone-Moncrief (FM) torus onto a spinning black hole. Rather, it has been observed to emerge in simulations that present considerable differences to the idealized FM setup traditionally employed in GRMHD simulations of accretion flows, provided that sufficient net poloidal magnetic flux accumulates near the black hole over the relevant accretion timescale. Specifically, \cite{ressler2021} simulated the spherical accretion of zero angular momentum matter with a uniform magnetic field onto spinning black holes, with their setup reaching states presenting several similarities to the MAD state achieved by typical FM setups. A long-duration study of a comparable Bondi-like feeding configuration by \cite{lalakos2024} showed that such a system can develop a more traditional MAD state with powerful jets once sufficient magnetic flux accumulates near the horizon, while subsequently exhibiting additional transitions in its accretion and jet behavior. Generalizing this setup to the spherical accretion of low angular momentum matter, \cite{galishnikova2025} found that systems of this kind also reach states with similarities to the typical MAD state over the simulated duration, albeit with weaker jets and stronger variability.

Moreover, using a multiscale GRMHD approach, \cite{guo2025} observed the emergence of the typical MAD state in simulations of accreting FM tori of a much larger scale than typical ones, while \cite{lalakos2025}, utilizing contiguous long-duration GRMHD simulations of accretion of an initially uniform medium onto a rotating black hole, observed the emergence of the MAD state for all scale separations considered, determined by the value of the Bondi radius. Taken together, these results indicate that MAD formation is possible beyond the traditional FM torus setup, but that the physical state of the system depends on the feeding conditions and on whether it has evolved for sufficiently long times to reach its asymptotic behavior. In this respect, these studies may correspond to different evolutionary stages of similar systems, while the standard FM torus setup largely bypasses the initial flux-accumulation timescale. The timescale required to reach a time-averaged MAD state may also depend on the angular-momentum content of the inflow.

In the MAD state, the excess magnetic flux in the vicinity of the black hole is episodically expelled back into the disk in what are known as flux eruption events\citep{igumenshchev2008, tchekhovskoy2011}. During these events, coherent, long-lived bundles of vertical magnetic field are generated close to the equatorial plane \citep{ripperda2020, ripperda2022}. These outwardly moving flux tubes are filled with low density, highly magnetized plasma, with their footpoints anchored in regions of the equatorial matter distribution where the vertical magnetic field is dynamically dominant \citep{porth2021}. This whole process repeats multiple times over the duration of a typical simulation in the MAD state, resulting in the high temporal variability exhibited both by the mass accretion rate and the magnetic flux through the black hole horizon in simulations of MADs \citep{tchekhovskoy2011, narayan2022, chatterjee2022}. 

Recent studies have emphasized that hotspots formed via reconnection are central to understanding the flaring activity observed in systems such as Sgr A* \citep{younsi2015, nathanail2022, lin2023, elmellah2023, dimitropoulos2025}. \cite{antonopoulou2025} showed that magnetic reconnection during flux eruption events in MADs produces hot spots that travel outward along newly formed magnetized flux tubes, giving rise to near-infrared flares whose properties match the observational characteristics of Sgr A* \citep{gravity2018}. 

Crucially, the same eruptive processes that drive hot spot formation also create favorable conditions for particle acceleration, namely the formation of magnetic reconnection sites \citep{sironi2014, werner2017}. Specifically, \cite{vos2024} found that reconnection during flux eruptions can efficiently accelerate particles of both hadronic and leptonic nature, and trigger enhanced pair production. These energetic particles and their secondary products contribute to the total cosmic-ray reservoir of a galaxy, which shapes the diffuse X-ray, $\gamma$-ray, and neutrino emission \citep{amenomori2021, kantzas2024} and can affect the physical and chemical properties of the interstellar medium on a galactic scale \citep{koutsoumpou2025a, koutsoumpou2025b}. 

Since reconnection is central to the dynamics of flux eruption events, resistivity of the accreting plasma can possess a critical role in shaping the evolution of these events and consequently the dynamical evolution and variability of MADs. \cite{nathanail2025} studied the effects of including physical resistivity in GRMHD simulations of MADs, finding that low resistivity models are, like their ideal MHD counterparts, highly variable, due to their evolution being determined by frequent reconnection events, while a moderate resistivity can effectively diffuse the accumulated magnetic flux. Moreover, \citep{aktar2025} corroborated the results of \citep{nathanail2025} and additionally showed that high-resistivity can efficiently diffuse the accreting matter's magnetic field, suppressing plasmoid formation and turbulence related to the magnetorotational instability (MRI), as well as leading to significantly lower jet powers.

In this work, we investigated the generation of vertical flux tubes and the evolution of the non-axisymmetric features of the equatorial accretion flow during a flux eruption event, by utilizing the results of a standard 3D GRMHD simulation which reaches the MAD state. In Sec. \ref{sec:2} we present the setup of our numerical simulation. In Sec. \ref{sec:3} we describe the mechanism via which vertical magnetic flux tubes are detached from the black hole horizon and subsequently transported outwards, effectively reducing the accumulated magnetic flux. The description of this mechanism is supported by results from our numerical simulation. Section \ref{sec:4} is dedicated to the examination of the azimuthal structure of the equatorial plasma distribution, featuring quantitative results regarding the evolution of the equatorial flow's non-axisymmetric features and dominant azimuthal modes, which are intrinsically connected to the formation of the vertical flux tube. We outline our conclusions in Sec. \ref{sec:5}.
		
\section{Numerical setup}\label{sec:2}
		
\subsection{Simulation parameters}
		
We utilize the results of a three-dimensional GRMHD simulation performed with the open source module of the MPI-AMRVAC framework, \textsc{BHAC} \citep{porth2017}. \textsc{BHAC} employs second order shock-capturing finite volume methods to solve the equations of ideal GRMHD
\begin{equation}
	\nabla_{\mu}(\rho u^{\mu}) = 0\, ,
\end{equation}
\begin{equation}
\nabla_{\nu}T^{\mu\nu} = 0\, ,
\end{equation}
\begin{equation}
	\nabla_{\mu}(\prescript{\star}{}{F}^{\mu\nu}) = 0\, ,
\end{equation}
where $\rho$ is the fluid's rest mass density, $u^{\mu}$ its four-velocity, $T^{\mu\nu}$ the energy-momentum tensor, and $\prescript{\star}{}{F}^{\mu\nu}$ the Hodge dual of the electromagnetic tensor, $F^{\mu\nu}$. The code also utilizes constraint transport methods compatible with adaptive mesh refinement (AMR) \citep{londrillo2004, delzanna2007} to satisfy the solenoidal condition, $\vec{\nabla}\cdot\vec{B} = 0$, for the magnetic field \citep{olivares2019}. \textsc{BHAC} has been widely used for numerical investigations of various relativistic environments and processes \citep{nathanail2019, cruz-osorio2022, mpisketzis2024, das2024} and has undergone extensive benchmarking against similar codes \citep{porth2019}.
		
The three-dimensional simulation was performed in modified Kerr-Schild (KS) coordinates \citep{mckinney2004} and features an effective resolution of  $N_{r}\times N_{\theta}\times N_{\phi} = 384  \times 192 \times 192$. As initial conditions, we used a standard, axisymmetric Fishbone-Moncrief (FM) torus \citep{fishbone1976} with a constant specific angular momentum of $l = 6.76$, in hydrodynamic equilibrium around a Kerr black hole with a dimensionless spin of $\alpha = J/M^{2} = 0.94$ and outer horizon radius of $r_{\mathrm{H}} \simeq 1.34\, r_{g}$. The inner edge of the initial torus was set at $r_{\text{in}} = 20\, r_{g}$ from the central black hole, while the location of the density maximum was located at $r_{\text{max}} = 40\, r_{g}$, where $r_{g} = M$ is the gravitational radius of the central Kerr black hole of mass $M$, in units in which $G = c = 1$. Lastly, we used an ideal equation of state with a constant adiabatic index, $\hat{\gamma} = 4/3$.
		
The initial magnetic field was purely poloidal and determined by the vector potential
\begin{equation}
A_{\phi} \propto \max\left[\dfrac{\rho}{\rho_{\text{max}}}\left(\dfrac{r}{r_{\text{in}}}\right)^{3}\mathrm{e}^{-r/r_{\text{mag}}}\sin^{3}{\theta} - A_{\phi, \text{cut}}, 0\right]\,
\end{equation}
where $A_{\phi,\text{cut}} = 0.2$, $r_{\text{mag}} = 400\, r_{g}$, and $\rho_{\text{max}} = 1$ is the initial density distribution's maximum value. This prescription for the initial magnetic vector potential corresponds to a purely poloidal magnetic field in a single nested loop configuration and is standard for simulations which reach the quasi-steady MAD state \citep{wong2021, zhang2024}. Furthermore, the strength of the initial magnetic field was normalized so that the ratio of the maximum gas pressure over the maximum magnetic pressure in the initial FM torus, $\beta_{\text{max}} = \dfrac{(P_{g})_{\text{max}}}{(P_{\text{mag}})_{\text{max}}} = 100$. We note that $(P_{g})_{\text{max}}$ and $(P_{\text{mag}})_{\text{max}}$ do not necessarily occur in the same location in the initial torus. Finally, numerical density floors are treated in the standard manner for GRMHD codes \citep{porth2017, nathanail2020}.
\begin{figure}
    \centering
    \includegraphics[width = 0.5\textwidth]{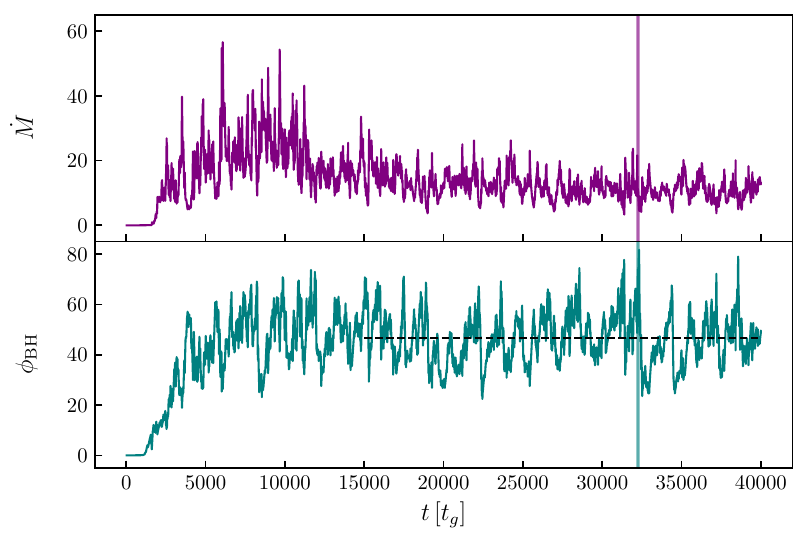}
	\caption{Time series of the mass accretion rate, (top panel) and normalized magnetic flux (bottom panel) on the black hole outer horizon for the entirety of our simulation The horizontal dashed black in the plot of $\phi_{\text{BH}}$ corresponds to its time average $\langle \phi_{\text{BH}} \rangle_{t} = 47$.}
	\label{mdotfig}
\end{figure}

\subsection{Identification of flux eruption events}
		
We define the mass accretion rate, $\dot{M}$, as the integral of the mass flux, $\rho u^{r}$, over the surface of the black hole's event horizon, which is located at $r \simeq r_{\mathrm{H}}$. This integral is
\begin{equation}\label{mdot}
	\dot{M}(t) = \int_{\phi}\int_{\theta}\rho u^{r}dA_{\theta\phi}\, .
\end{equation}
Similarly, the dimensional or absolute magnetic flux at the event horizon, $\Phi$, is
\begin{equation}\label{phibh}
	\Phi(t) = \dfrac{\sqrt{4\pi}}{2}\int_{\phi}\int_{\theta}\left|B^{r}\right|dA_{\theta\phi}\, .
\end{equation}
 Following convention, we have multiplied the magnitude of the radial magnetic field, $\left|B^{r}\right|$ with $\sqrt{4\pi}$, in order to express it in Gaussian units. Additionally, the factor $1/2$ in Eq. \ref{phibh} appears in order to specify that $\Phi$ is calculated over one hemisphere. The dimensionless normalized magnetic flux, $\phi_{\text{BH}}$, can be calculated from $\dot{M}$ and $\Phi$ as
\begin{equation}
	\phi_{\text{BH}}(t) = \dfrac{\Phi(t)}{\sqrt{\dot{M}(t)}}\, .
\end{equation}
In this system of units, the normalized horizon magnetic flux saturation value is $\phi_{\text{BH}} \simeq 50$ \citep{tchekhovskoy2011}. The mass accretion rate, $\dot{M}$ and normalized magnetic flux, $\phi_{\text{BH}}$, at the black hole horizon over the duration of our simulation are presented in Fig. \ref{mdotfig}.
		
After $t \simeq 12000\, t_{g}$, the simulation has reached the MAD state and $\dot{M}$, $\phi_{\text{BH}}$ display significant temporal variability, typical for accretion in the MAD state. At certain times in the simulation, $\phi_{\text{BH}}$ reaches a value much higher than its saturation value, with the mass accretion rate plummeting at the same time. During these phases, the absolute magnetic flux, $\Phi$, displays a decrease from a local maximum value. This suggests the removal or expulsion of magnetic flux threading the event horizon. Consequently, we consider time intervals over which $\Phi$ is initially at a local maximum and then decreases while $\phi_{\text{BH}}$ surpasses its saturation value as corresponding to flux eruption events.

\begin{figure}
	\centering
	\includegraphics[width = 0.5\textwidth]{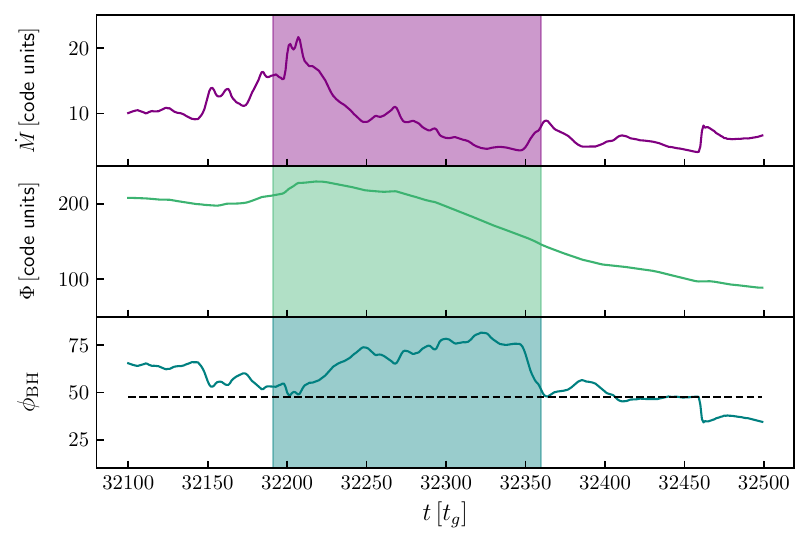}
	\caption{Mass accretion rate, $\dot{M}$, (top panel), dimensional magnetic flux, $\Phi$ (middle panel), and normalized magnetic flux, $\phi_{\text{BH}}$, (bottom panel) during the flux eruption event at $t = 32220\, t_{g}$. The shaded areas of each plot denote the time span of interest.}
	\label{eruption}%
\end{figure}
		
We identify multiple such events during our simulation and focus our analysis on an event observed over the time interval $\Delta t = 32191\text{--}32360\, t_{g}$, with the onset of the event occurring at $t = 32220\, t_{g}$, as marked in the time series presented in Fig.~\ref{eruption}. We also present results related to three additional events over the analysis intervals $\Delta t = 14990 \text{--} 15290\, t_{g}$, $\Delta t = 21890 \text{--} 22390\, t_{g}$, and $\Delta t = 34190 \text{--} 34300\, t_{g}$ in Appendix~\ref{appendixA}. For these three events the start of the respective analysis intervals corresponds to the onset of the flux eruption. The onset time reported for each event is defined as the time at which the absolute magnetic flux threading the horizon reaches a local maximum within the corresponding analysis interval.

During the main event, at $t = 32200\, t_{g}$, $\dot{M}$ and $\Phi$ both display a short increase, reaching their respective local maxima at $t = 32207\, t_{g}$ and $t = 32220\, t_{g}$. Both quantities subsequently decrease, while the normalized horizon magnetic flux, $\phi_{\text{BH}}$, obtains values higher than its saturation value over the entirety of the time interval of interest. We note that, due to the strong and rapid variability of the accretion flow in the MAD state and especially during flux eruption events, the extent of the radial inflow equilibrium region also shows significant variations. On average, after the system has achieved its quasi-steady state, the radial inflow equilibrium region extends out to $r \geq 40\, r_{g}$.
		
\section{Formation and motion of vertical flux tubes}\label{sec:3}
		
Accreting systems in the MAD state are characterized by magnetic flux eruptions, episodic events during which magnetic flux in the vicinity of the black hole is expelled outwards \citep{porth2021}. During these events, magnetic reconnection forms hotspots in the equatorial accretion disk, as well as vertical magnetic flux tubes which thread the disk \citep{ripperda2020, ripperda2022}. A deep understanding of the mechanism which triggers flux eruption events and leads to the formation of their characteristic accompanying structures is necessitated by their connection to the observed flaring behavior of accreting environments \citep{dexter2020, hakobyan2023, antonopoulou2025}. 
		
\subsection{The detachment instability mechanism}
		
\begin{figure*}
	\centering
	\includegraphics[width = 1.0\textwidth]{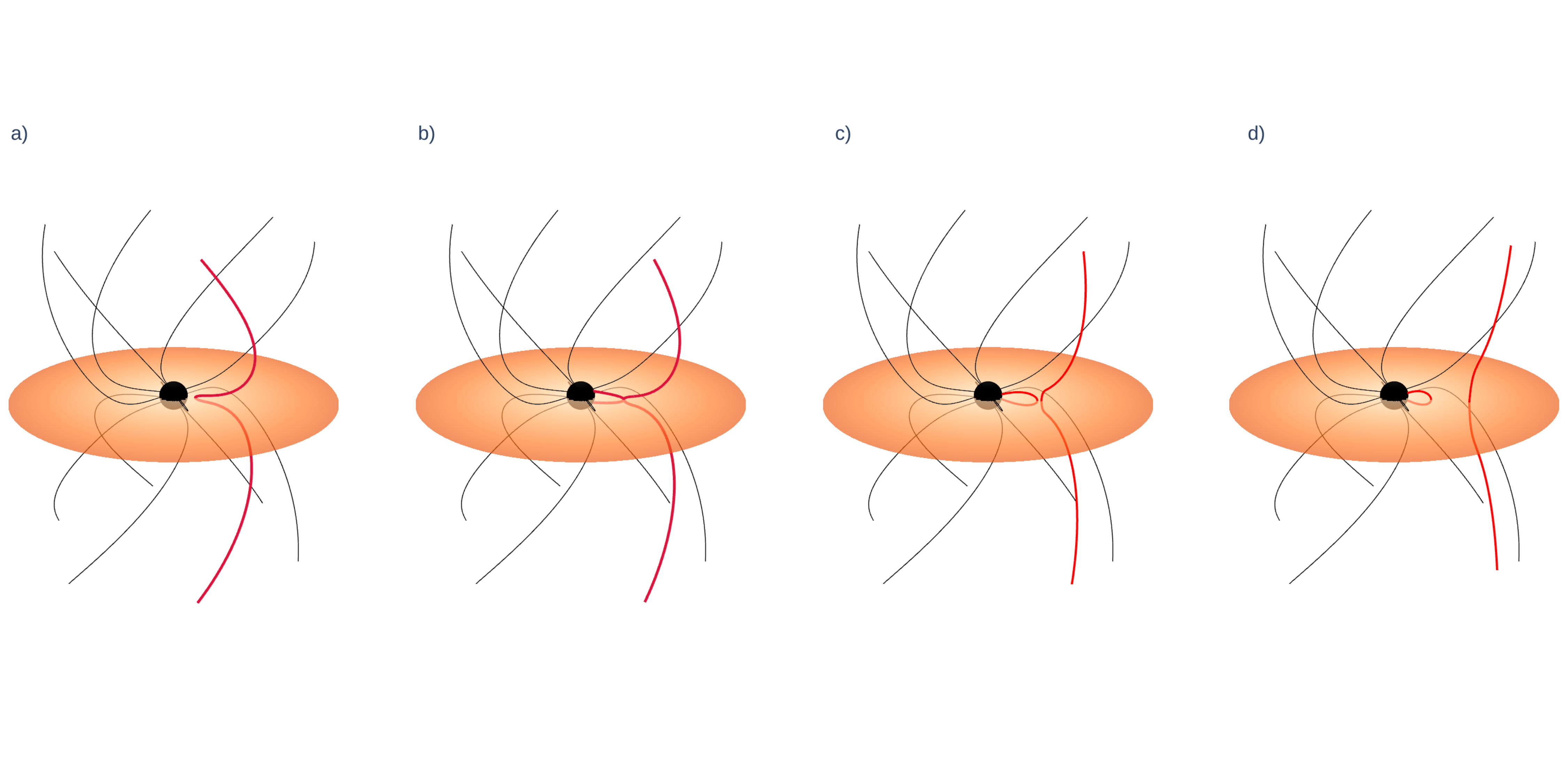}
	\caption{Schematic depiction of the physical mechanism through which magnetic flux is expelled from the black hole. Magnetic field lines threading the horizon are shown as thin black lines. The bundle of magnetic field lines of interest is depicted as a thick red line. The temporal evolution of the mechanism is shown from left to right: a) Magnetic flux is advected by the accreting matter towards the black hole. Close to the black hole the magnetic field is mostly equatorial. b) The equatorial magnetic field slightly above and below the midplane reconnect. c) Two structures form during the magnetic reconnection of the equatorial magnetic field, a magnetic field loop close to the black hole's event horizon, and a tube of vertical magnetic field lines. d) The vertical flux tube, filled with plasma of lower density than its environment, is buoyantly carried away from the black hole.}
	\label{sketch}%
\end{figure*}

Magnetic reconnection possesses a central role in the formation of the vertical flux tubes associated with flux eruption events, by essentially reconfiguring the equatorial magnetic field on the accretion disk into bundles of vertical magnetic field lines \citep{ripperda2020, ripperda2022}. These vertical magnetic flux tubes are large scale structures comprised of coherent vertical magnetic field lines, filled with low-density plasma \citep{porth2021, salas2024}. The generation and subsequent motion of the vertical flux tubes is a physically rich and complex multistage process, which we summarize schematically in Fig. \ref{sketch}.
		
At the onset of the flux eruption event, the disk magnetic field close to the black hole is dominated by its horizontal component, i.e. its component parallel to the midplane. Magnetic field is advected along with the accreting matter, also in a horizontal configuration at small radii (panel a of Fig. \ref{sketch}). As mass slides through the field lines connected to the black hole, the upper parts of the flux tube become  magnetically buoyant and push outwards away from the black hole, via a form of the Parker instability \citep{parker1955, parker1966}. This increases the tension of the magnetic field lines. The tension increases further as matter pushes the field lines closer to the black hole, and the field lines eventually become tightly compressed above and below the equator. 
		
At some point, two horizontal magnetic field lines close to the black hole, one above and one below the midplane, reconnect, forming an X-point on the equatorial plane, typically at a distance of a few (3-5) $r_{g}$ from the black hole (panel b of Fig. \ref{sketch}). The reconnection of the two horizontal magnetic field lines leads to the formation of two structures, a magnetic field loop close to the black hole's horizon, and a vertical  magnetic flux tube which threads the equatorial plane (panel c of Fig. \ref{sketch}). 
Part of the reconnecting field is accreted with the plasma, leading to the detachment of the flux tube (panel d of Fig. \ref{sketch}). Neighboring magnetic field lines immediately follow, with the reconnecting region's extent on the midplane expanding. Thus, magnetic field lines are continuously detached from the black hole and added to the flux tube as part of a cascading process, which we term detachment instability. The vertical flux tube, filled with plasma of lower density than its surrounding environment, is magnetically buoyant and transported outwards away from the black hole. Through this process, the excess magnetic flux concentrated in the vicinity of the horizon is detached from the black hole and then buoyantly transported away, resulting in the observed decrease of the horizon magnetic flux during flux eruption events.

The detachment instability described in this subsection is a mechanism wholly distinct from the typical MHD instabilities, such as the magnetic Rayleigh-Taylor or the Kelvin-Helmholtz instability that may emerge in a turbulent plasma flow, like an accretion disk. It is not driven by a tendency of heavy fluid parcels to decrease their potential energy as in the case of the magnetic Rayleigh-Taylor instability, or by velocity shear between adjacent layers of different characteristics, as in the case of the Kelvin-Helmholtz instability, and does not lead to mixing like these two characteristic MHD instabilities. Instead, the detachment instability is a reconnection-driven process. It arises in the low-density, highly magnetized inner disk where horizontal fieldlines above and below the midplane reconnect, reconfiguring the topology of horizon-threading flux lines into vertical bundles. Subsequently, neighboring field lines follow, producing a cascade of reconnection-induced reconfiguration in which magnetic flux is continuously removed from the black hole and added to the growing flux tube, after which the tube, being strongly magnetized and less dense than its environment, is buoyantly transported outward.
		
\subsection{Numerical results}
		
The physical scenario described in the previous subsection is supported by the results of our standard 3D GRMHD simulation. As explained in the previous section, we identified multiple flux eruption events during our simulation, focusing our analysis on an event starting at $t = 32220\, t_{g}$. In order to study in detail the conditions in the equatorial accretion flow during this flux eruption, we obtained outputs with a frequency of $1\, t_{g}$. We repeated our analysis for three additional events, one at $t = 14990\, t_{g}$, one at $t = 21890\, t_{g}$, and one at $t = 34190\, t_{g}$, for results with a lower output frequency of $50\, t_{g}$. These results, as well as a convergence check of the results of the flux eruption event of the main text with the output frequency, are presented in Appendix \ref{appendixA}.
		
\begin{figure}
	\centering
	\includegraphics[width = 0.5\textwidth]{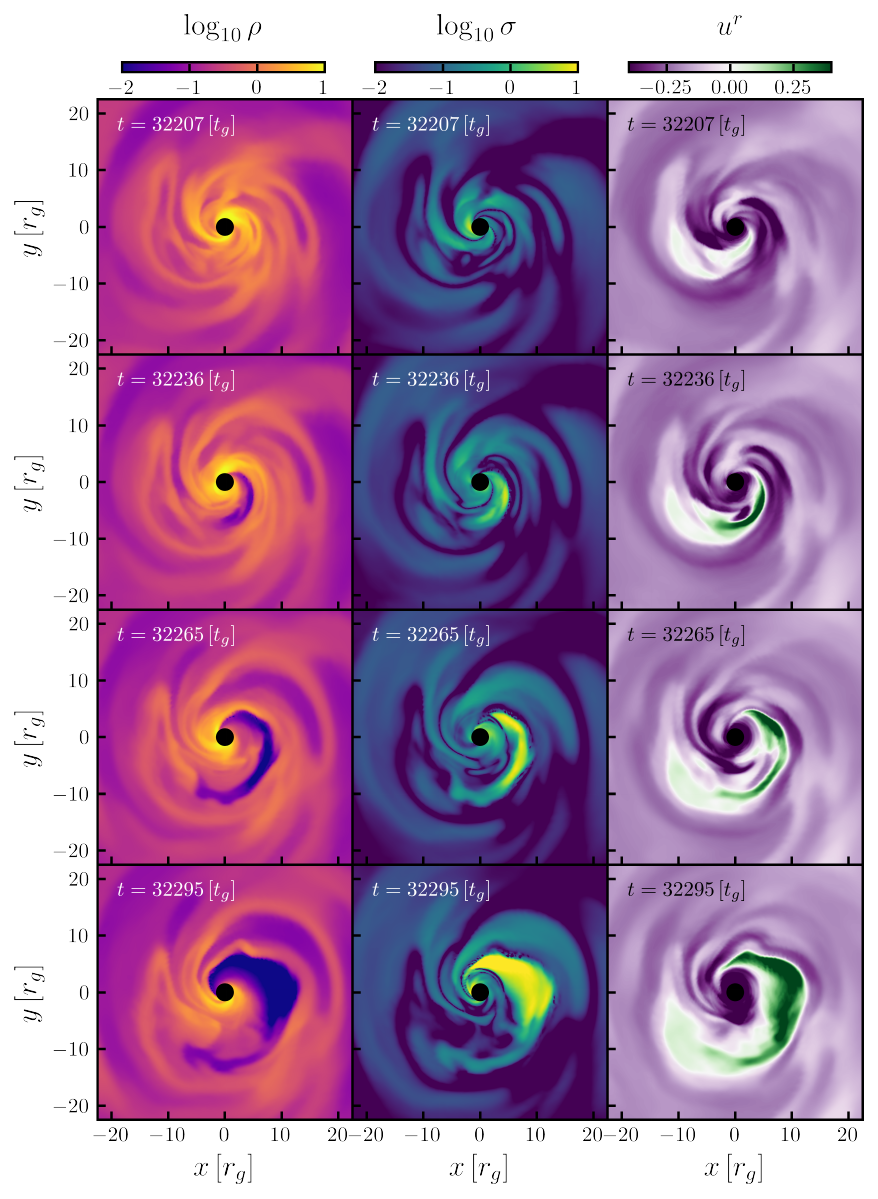}
	\caption{Equatorial plasma density, $\rho$ (left), magnetization, $\sigma$ (middle), and radial four-velocity $u^{r}$ (right) at four different snapshots during the flux eruption event. The first row corresponds to the maximum of $\dot{M}$ at $t = 32207\, t_{g}$. An extended region of high magnetization and low density forms during the event. The low-density plasma in the high-magnetization region has a positive radial four-velocity. The plasma at the footpoint of the flux tube, which begins forming at the edge of the high $\sigma$ region at negative $y$, also has positive radial four-velocity.}
	\label{eq1}%
\end{figure}

We first turn our attention to the distribution and kinematics of the accreting plasma on the equatorial plane, by analyzing equatorial slices of the plasma density, $\rho$, magnetization, $\sigma = b^{2}/\rho$, and radial four-velocity, $u^{r}$. These quantities indicate the gradual formation of an expanding region of low density, where the plasma is strongly magnetized and moving outwards with a strongly positive radial velocity, as shown in Fig. \ref{eq1}. 
		
\begin{figure}
	\centering
	\includegraphics[width = 0.5\textwidth]{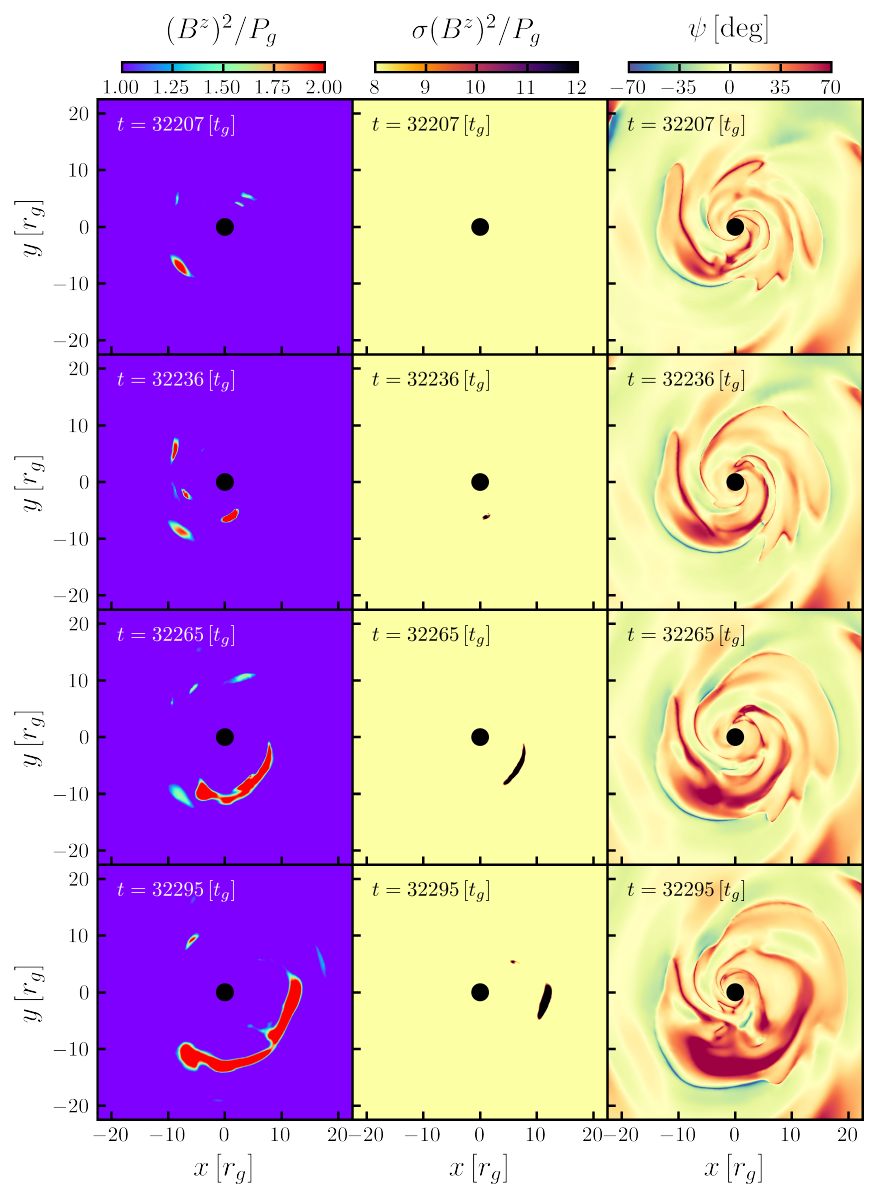}
	\caption{$(B^{z})^{2}/P_{g}$ (left), $\sigma(B^{z})^{2}/P_{g}$ (middle), and angle $\psi$ (right) of the magnetic field with the equatorial plane. At $t = 32236\, t_{g}$ a bundle of ordered vertical magnetic field lines has began forming, with its most energetic part threading the midplane at a distance of about $6\, r_{g}$ from the black hole. The area of the flux tube footpoint expands over time as more equatorial magnetic field lines reconnect and form additional vertical magnetic lines, seen in the middle column.}
	\label{eq2}%
\end{figure}

The area of the equatorial plane corresponding to the footpoint of the vertical flux tube that is generated during the flux eruption is better identified by considering quantities related to the magnetic field. The first of these quantities is defined as $(B^{z})^{2}/P_{g}$, with $(B^{z})^{2}/P_{g} \geq 1$ in regions where the vertical magnetic field dynamically dominates over the matter \citep{porth2021}. The second quantity related to the magnetic field is $\sigma (B^{z})^{2}/P_{g}$. This quantity traces the most energetic part of the flux tube footpoint \citep{antonopoulou2025}, i.e. the region of the midplane of newly formed vertical magnetic field by magnetic reconnection. The third quantity related to the magnetic field is the angle, $\psi$, between the magnetic field and the equatorial plane, defined as
\begin{equation}
	\psi = \arctan\left(\dfrac{|B^{z}|}{|B_{\text{hor}}|}\right)\, ,
\end{equation}
where $B_{\text{hor}}$ is the horizontal component of the magnetic field. These three quantities are displayed in Fig. \ref{eq2}. 
		
Initially, there is only a small region of dominant magnetic field close to the black hole, which has formed before the onset of the event and slowly disappears during the event. This region is visible in the first three snapshots of the left column of Fig. \ref{eq2}. In the second snapshot, shown in the left column of Fig. \ref{eq2}, a region where a strong $B^{z}$ threads the disk has began forming. This region is the footpoint of the vertical flux tube which is generated during the event. That same region at the same timestamp is also characterized by a high value of $\sigma(B^{z})^{2}/P_{g}$, which traces regions of newly reconnected field lines on the midplane. In the next timestamp, after $29\, t_{g}$, this region has expanded, as more horizontal field lines reconnect into a vertical configuration. The black region of high $\sigma(B^{z})^{2}/P_{g}$ in the middle column of Fig. \ref{eq2}, has a smaller extent than the region of strong $B^{z}$ depicted in the left column of Fig. \ref{eq2}, however, it much more accurately traces the most energetic part of the flux tube's footpoint. We also note that this region is located at the edge of the low-density and high-magnetization region shown in Fig. \ref{eq1}. 
		
The angle $\psi$ of the magnetic field with the equatorial plane shows that the magnetic field is mostly horizontal in the high-magnetization region, having a strong vertical component at the edge of this region. The configuration of the magnetic field at the edge of the expanding high-magnetization region is highly transient. However, in the large region of $\psi \geq 70 \, \mathrm{deg}$ located at negative $y$ at late times of our time interval, the dominant vertical magnetic field has achieved a slowly evolving and long-lasting configuration. This region corresponds to the footpoint of the vertical flux tube which has formed during the event. This region is filled with plasma of lower density than that of the surrounding medium. As such, it slowly moves away from the black hole as a result of magnetic buoyancy, as shown by its positive radial four-velocity in Fig. \ref{eq1}. This buoyant outwards motion of the vertical flux tube formed during our analysis time interval holds similarities with the motion of buoyant, low-density flux tubes which transport magnetic flux away from the magnetospheric boundary in instances of stellar accretion \citep{takasao2018, takasao2022}. Interestingly, this reconnection-based flaring mechanism for the ejection of magnetic flux from the central black hole as a whole, has been observed in 3D simulations of accretion onto magnetized protostars \citep{takasao2019}. 

In fact, the similarities between accretion onto protostars and the MAD regime of black hole accretion are remarkable. \cite{takasao2019} performed simulations of accretion onto protostars lacking a magnetosphere of their own. They found that the accreting material leads to the accumulation of magnetic flux onto the protostar, forming a magnetospheric boundary after accretion initiates. Equatorial field lines of opposite polarity reconnect close to the midplane, leading to the development of areas of low density and high magnetization within the disk which are ejected from the accreting system. This mechanism they invoke for the generation of protostellar flares is similar to the detachment instability mechanism, indicating a common regime of accretion onto objects that do not posses an ab-initio magnetospheric boundary but instead build one up after accretion commences due to the accumulation of magnetic flux dragged in by the accreting plasma. In this common regime of accretion, reconnection at and around the equatorial plane determines the morphology and evolution of the inner accretion flow. This is different to accretion onto objects possessing a strong magnetosphere of their own, where the magnetic Rayleigh-Taylor instabilities at the magnetospheric boundary determine the inner accretion flow dynamics and morphology \citep{kulkarni2008, kulkarni2009, blinova2016, parfrey2024}. 
		
\section{Azimuthal structure of the equatorial accretion flow}\label{sec:4}
		
The results presented in the previous section indicate that the distribution of the accreting matter on the equatorial plane changes significantly over the duration of the flux eruption event. Specifically, the equatorial accretion flow becomes increasingly less axisymmetric over the duration of the flux eruption. The non-axisymmetric features which emerge during the eruption event are of major importance to the dynamical evolution of the accretion flow, as they enable the infalling plasma to overcome the magnetic barrier close to the black hole \citep{narayan2003, igumenshchev2008, mckinney2012, porth2021}.
		
\subsection{Measuring the accretion disk's degree of axisymmetry}
		
Numerical simulations of accretion onto black holes have shown that the emergence of magnetically dominated low-density regions and spiral patterns of infalling matter on the equatorial plane are a feature that always appears during the evolution of accreting systems in the MAD state \citep{igumenshchev2008, tchekhovskoy2011, porth2021, ripperda2022}. The emergence of these features disrupt the axisymmetry of the equatorial accretion flow and shape the azimuthal structure of the equatorial matter distribution. In this subsection, we present a simple quantitative measure of the degree of non-axisymmetry exhibited by the equatorial matter distribution, based on the Fourier expansion of the mass density.
		
The Fourier expansion of the equatorial mass density constitutes a robust method for deciphering the azimuthal structure of the matter distribution on the midplane. A similar method based on Fourier expansions has been utilized in past works for uncovering the azimuthal structure of standing shock surfaces in simulations of accretion flows \citep{nagakura2008, garain2023}.
		
We consider the density $\rho(t,\phi)|_{r}$ along the circumference of a circle of radius $r$ on the equatorial plane. At each snapshot in our analysis interval, this density can be expressed as a Fourier series with respect to $\phi$ (equivalently a trigonometric series). The coefficients of this series can be calculated as
\begin{equation}\label{series}
	D_{m}(t) = \dfrac{1}{2\pi}\int_{0}^{2\pi}\rho(t,\phi)|_{r}\mathrm{e}^{-i m\phi}d\phi\, .
\end{equation}
For $m=0$, the above relation returns the average or axisymmetric term in the density Fourier series, with higher order terms giving the coefficients which express the non-axisymmetric corrections to the axisymmetric term. The Fourier series of the density is calculated up to $N = 200$. This ensures that the series is fully converged to the resolution determined by the simulation grid, which is $N_\phi = 192$. We employ the  standard criterion for GRMHD simulations, which requires the wavelength of the fastest-growing mode of the MRI to be resolved by a minimum of 6 grid cells in the azimuthal direction. With an azimuthal resolution of $N_{\phi} = 192$, the largest azimuthal mode number that can be reliably resolved is $m_{\text{max}} = \dfrac{N_\phi}{6} = \dfrac{192}{6} = 32$. Therefore, structures at $m > 32$ are cannot be reliably resolved.

\begin{figure}
	\centering
	\includegraphics[width = 0.5\textwidth]{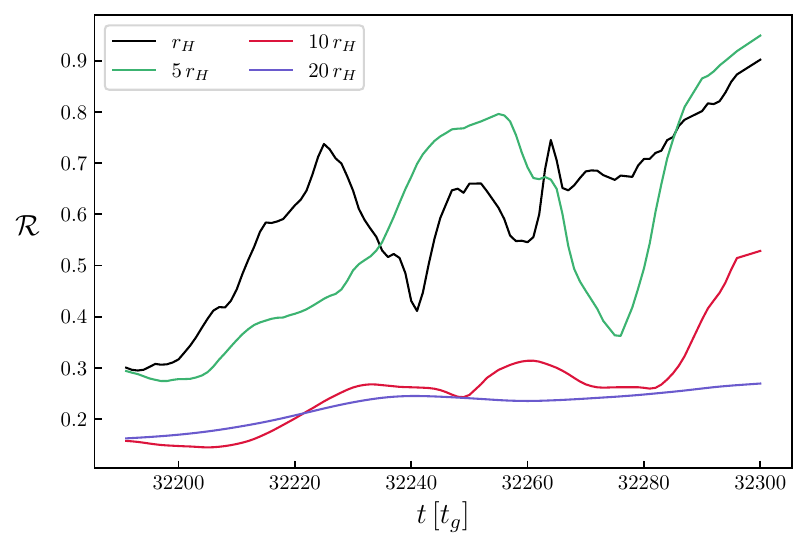}
	\caption{Measure of the equatorial matter distribution's degree of non-axisymmetry $\mathcal{R}$ at four different radial positions over the time interval of interest.}
	\label{axisym}%
\end{figure}

Additionally, we define the measure of the degree of non-axisymmetry exhibited by the equatorial matter distribution at each timestamp at a specific radius $r$ as
\begin{equation}
	\mathcal{R}(t) = \dfrac{2\sum_{m = 1}^{N} |D_{m}(t)|^{2}}{|D_{0}(t)|^{2}}\, .
\end{equation}
$\mathcal{R}(t)$ measures the relevant contribution of the non-axisymmetric ($\sum_{m = 1}^{N} |D_{m}(t)|^{2}$) and axisymmetric ($|D_{0}(t)|^{2}$) components to the Fourier series, and as such is a suitable metric for the degree of non-axisymmetry displayed by the equatorial matter distribution. In Fig. \ref{axisym}  we present $\mathcal{R}$, calculated at every snapshot of our time interval and at radii $r = \{r_{\mathrm{H}}, 5\, r_{\mathrm{H}}, 10\, r_{\mathrm{H}}, 20\, r_{\mathrm{H}}\}$, which shows how the accretion flow's morphology transitions from almost axisymmetric to highly non-axisymmetric. 
		
At all radii that we considered in our analysis, the equatorial matter distribution displays a strengthening of its non-axisymmetric features, with this effect being more prominent close to the black hole, at $r \simeq r_{\mathrm{H}}$ and $r = 5\, r_{\mathrm{H}}$. At larger distances from the black hole, the amplification of the non-axisymmetric features of the inner accretion flow is considerably less significant. Moreover, $\mathcal{R}$ displays stronger variability closer to the black hole, thus revealing the transient nature of the structures which form in the innermost regions of the accretion disk. The enhancement of large scale non-axisymmetric features during the flux eruption event, is much more significant in the inner parts of the equatorial accretion flow.

\begin{figure*}
	\centering
	\includegraphics[width = 1.0\textwidth]{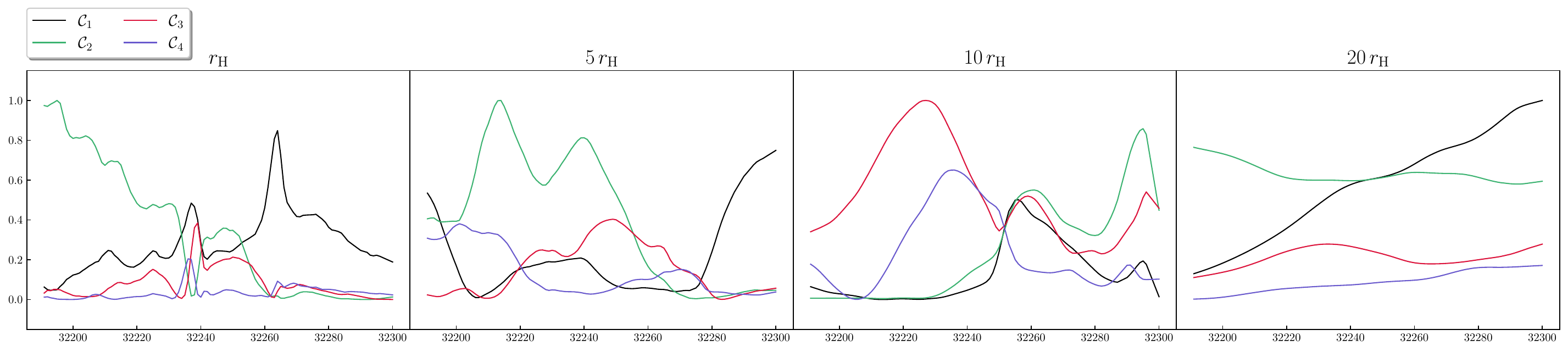}
	\caption{Relative contribution of the first three non-axisymmetric terms at all four radii considered in our analysis.}
	\label{ratios}%
\end{figure*}
        
We also examined the relative contribution of each of the first three non-axisymmetric terms ($m = 1-4$) in the expansion of $\rho$, through the ratios
\begin{equation}
	\mathcal{C}_{m}(t) = \dfrac{|D_{m}(t)|^{2}}{\max({|D_{m}(t)|^{2}})}\, ,
\end{equation}
presented in Fig. \ref{ratios}. Along the $r \simeq r_{\mathrm{H}}$ circumference, the initial increase in $\mathcal{R}$ is mainly a result of the amplification of the $m = 2$ terms of the Fourier series, while its further steady increase at later times in our time interval can be mainly attributed to the $m = 1$ term. At $r = 5\, r_{\mathrm{H}}$, the main non-axisymmetric features in the middle of the time interval appear due to the $m=1, 2$ terms of the series. At $r = 10\, r_{\mathrm{H}}$, the dominant non-axisymmetric contribution appears to be due to the $m = 3$ term, while at $r = 20\, r_{\mathrm{H}}$ the $m = 2$ azimuthal mode dominates over the first half of the time interval, before being surpassed by the $m = 1$ term. We note that the strong increase in $\mathcal{R}$ at $r = r_{\mathrm{H}}$ and $5\, r_{\mathrm{H}}$ at the end of the time interval is a result of the amplification of the $m = 1$ term.
		
\subsection{Dominant azimuthal modes}
\begin{figure}
	\centering
	\includegraphics[width = 0.5\textwidth]{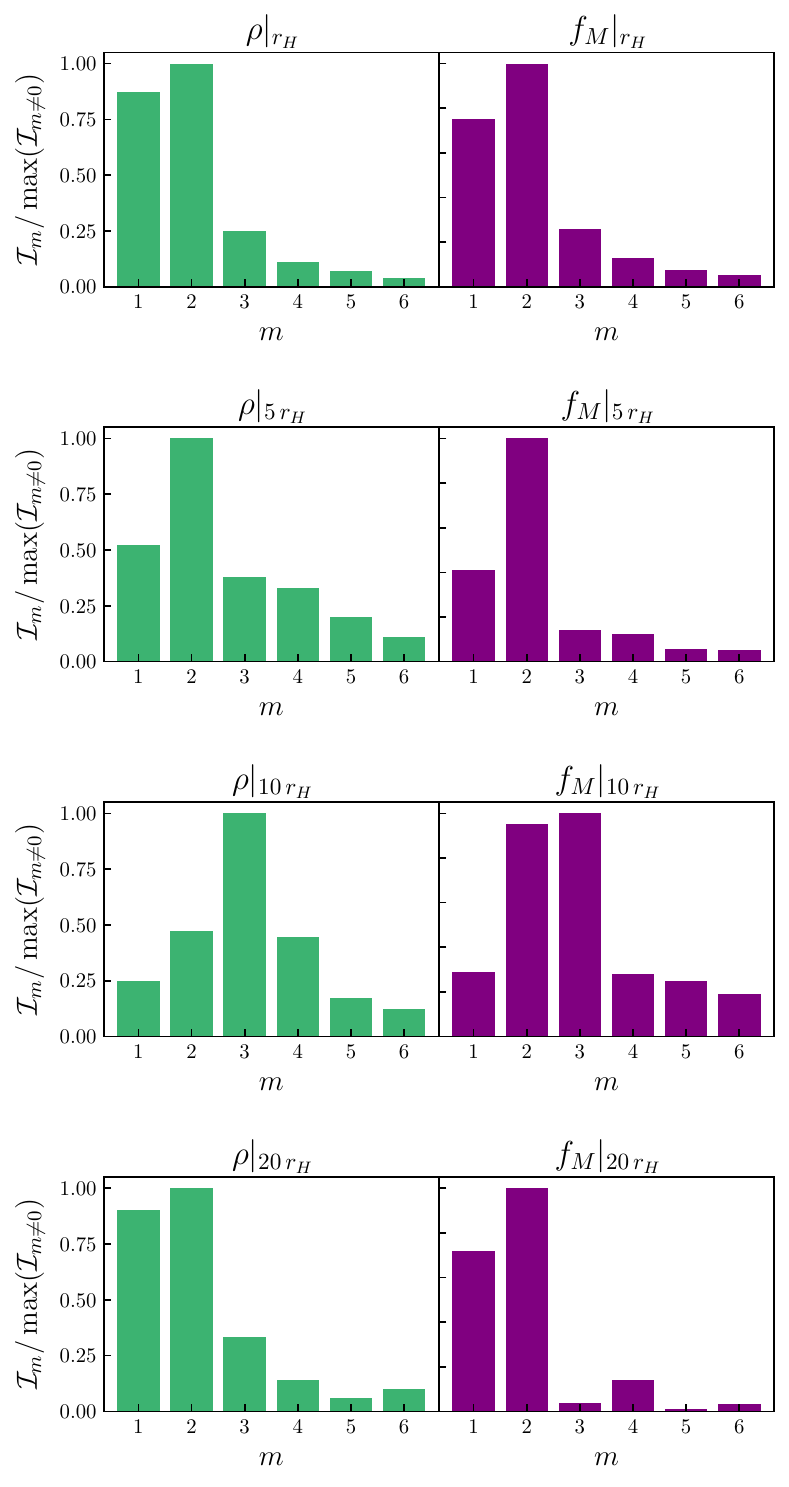}
	\caption{Time integrated squared Fourier coefficients, which express the power in each term of the expansion, for the equatorial plasma density, $\rho(t,\phi)$ and mass accretion rate per solid angle, $f_{M}(t,\phi)$, integrated over the entire time interval from $t = 32191\, t_{g}$ to $t = 32360\, t_{g}$. For the main flux-eruption event analyzed in the text, the integral corresponding to the $m = 2$ azimuthal mode is dominant at all radii except $r = 10\, r_{\mathrm{H}}$, where the $m = 3$ mode dominates.}
	\label{peaks}%
\end{figure}

We subsequently utilized the method based on the Fourier expansion of the equatorial plasma density introduced in Sec. \ref{sec:2} to determine the dominant azimuthal modes of fundamental quantities related to the equatorial matter distribution. Namely, additionally to the plasma density, we calculated the Fourier expansion coefficients of the mass accretion rate per solid angle, defined as
\begin{equation}
	f_{M}(t,\phi)|_{r} = \rho u^{r}\sqrt{-g}|_{r,\theta=\pi/2}\, .
\end{equation}
Subsequently, we calculated the integrals of the powers of each term in the expansions
\begin{equation}
	\mathcal{I}_{m} = \int|D_{m}(t)|^{2}dt\, ,
\end{equation}
over the entirety of the time interval. We performed these calculations for $\rho$ and $f_{M}$ at $r = \{r_{\mathrm{H}},\, 5\, r_{\mathrm{H}}, 10\, r_{H},\, 20\, r_{H}\}$. The results are shown in Fig. \ref{peaks}.
		
As observed in Fig. \ref{peaks}, the dominant azimuthal mode of the expansions of $\rho(t,\phi)|_{r_{\mathrm{H}}}$ and $f_{M}(t,\phi)|_{r_{\mathrm{H}}}$ for the main event is the $m = 2$ mode, with the $m = 1$ being a close second. The same behavior is observed at $r = 5\, r_{\mathrm{H}}$, where once again the $m=2$ mode is the dominant one. However, the relative contribution of the $m=1$ mode is lower compared to $r=r_{\mathrm{H}}$. Further away from the black hole, at $r = 10\, r_{\mathrm{H}}$, the $m = 3$ mode dominates the non-axisymmetric spectrum, before the $m = 2$ becomes the dominant mode again at $r = 20\, r_{\mathrm{H}}$. It is interesting to note that our results for the main event indicate the dominance of low azimuthal number modes on the equatorial plane, mainly of the $m = 1$ and $m = 2$ modes close to the black hole. Repeating this analysis for the three additional events at $t = 14990\, t_{g}$, $t = 21890\, t_{g}$, and $t = 34190\, t_{g}$ yields the same qualitative picture that the equatorial inner accretion flow is dominated by low-order azimuthal modes, although the specific mode with the largest time-integrated Fourier power varies between events and radii (see Appendix \ref{appendixA}). This indicates that the prominence of low-$m$ modes, especially $m = 1$ and $m = 2$, is a robust feature of the equatorial inner accretion flow during flux eruptions in the MAD regime. We note that a highly similar picture is observed in three-dimensional simulations of non-relativistic accretion onto magnetized stars, where accreting matter pierces its way through the central object's pre-existing magnetospheric boundary thanks to the development of low $m$ non-axisymmetric modes \citep{kulkarni2008, kulkarni2009, blinova2016, takasao2022, zhu2024}, even though in the MAD scenario the emergence of low $m$ non-axisymmetric modes is reconnection-induced, as in protostellar accretion \citep{takasao2019}, rather than MHD instability-driven, as in the aforementioned works.

We must also note that the angular momentum of the initial torus can have significant impact on the dynamical evolution and morphology of the accreting system during a flux eruption event. Specifically, in simulations with low angular momentum tori, magnetic flux tubes generated during flux eruption events are able to freely and quickly escape the inner disk due to magnetic buoyancy. On the other hand, strong azimuthal velocity shear in high angular momentum tori stretches the footpoints of magnetic flux tubes within the disks and leads to stronger mixing of the highly magnetized plasma within flux tubes with their weakly magnetized environment due to velocity shear, as shown recently by \citep{chan2025}. As such, the angular momentum of the initial torus is a parameter that can affect which azimuthal modes dominate the disk structure. 

Moreover, \cite{mckinney2012} report a dominance of the $m = 1$ mode close to the black hole in the density's Fourier expansion. This difference with respect to our results may arise from a combination of methodological and physical factors. On the methodological side, their Fourier analysis is performed on the density averaged over the full $\theta$ extent of the disk, whereas our analysis is based on equatorial slices, which probe a different aspect of the flow morphology. On the physical side, differences in the initial torus setup and in the resulting flow dynamics, including the rotation profile and associated shear, may also affect which azimuthal modes become most prominent. This discrepancy can therefore be attributed to a combination of factors relating to the analysis of simulation results and its physical setup.
		
The inner radii of an accretion flow in the MAD state constitute a region where interchange instabilities can develop and be intensified during energetic phases, such as the flux eruption events studied in this work. Firstly, the interface between the highly magnetized low-density region and its environment, i.e. the boundary of the flux tube footpoint, are areas where interchange instabilities develop. The differential rotation and local velocity shear of the inner accretion disk is conducive to the emergence of the Kelvin-Helmholtz instability, which affects the outer surface of the flux tube \citep{porth2021, ripperda2022}. Additionally, the inner parts of the accretion disk in the MAD state are strongly sub-Keplerian \citep{porth2021, dhruv2025}. This results in the plasma there experiencing a negative effective gravity, which is a favorable condition for the emergence of the magnetic Rayleigh-Taylor instability \citep{chandrasekhar1961}.
		
While these types of interchange instabilities develop in MADs, they most probably affect the disk at smaller scales, with the structures which form due to them possessing smaller angular extents and surviving for shorter timescales. Such smaller-angular-scale structures correspond to higher-$m$ modes, so their presence is most naturally interpreted as a signature of local fragmentation, shearing, and interchange activity in the inner flow, rather than of the large-scale morphology imposed by the reconnection-driven detachment instability.
		
\section{Conclusions}\label{sec:5}
		
In this work, we studied the configuration and dynamical evolution of the equatorial inner accretion flow around a rapidly spinning black hole during magnetic flux eruption events. We utilized the results of a 3D GRMHD simulation over a time interval which encompassed the duration of a flux eruption, with an output frequency of $1\, t_{g}$, in order to obtain a high temporal resolution of the transient dynamics of the inner accretion flow, and repeated our analysis for three additional flux eruption events, utilizing results with a lower output frequency of $50\, t_{g}$. Our key findings are summarized in the following paragraphs.
		
During the flux eruption, a bundle of vertical magnetic flux forms, threading the inner equatorial accretion flow. The formation of this structure is initiated by magnetic reconnection of horizontally oriented magnetic field lines, which have been advected inward by the accretion flow. As the magnetic field near the black hole becomes increasingly compressed and dynamically dominant, oppositely directed horizontal field lines above and below the equatorial plane reconnect, generating an X-point configuration close to the midplane. This reconnection event detaches a portion of the magnetic field from the horizon, reorganizing it into a vertical configuration threading the disk at a few gravitational radii. The resulting vertical flux tube, filled with plasma of lower density than its surroundings, becomes buoyantly unstable and rises outward, carrying magnetic flux away from the black hole. This mechanism enables the removal of excess magnetic flux from the horizon region and plays a fundamental role in regulating the saturation of the MAD state.
		
The flux eruption event is accompanied by the amplification of non-axisymmetric features in the equatorial accretion flow. As the magnetic field accumulates and reconnects, magnetically dominated low-density regions and dense accretion streams emerge on the midplane, breaking the initial axisymmetry of the disk. We quantified the evolution of the azimuthal structure by performing a Fourier decomposition of the equatorial mass density, computing the degree of non-axisymmetry at various radii. Our results show that the non-axisymmetric features are significantly enhanced closer to the black hole, particularly at $r \simeq r_{\mathrm{H}}$ and $r = 5\, r_{\mathrm{H}}$, while the outer disk remains comparatively more axisymmetric. The evolution of the non-axisymmetric features highlights the highly dynamic and transient nature of the inner accretion flow during magnetic flux eruptions in the MAD state.
		
We further analyzed the azimuthal structure of the equatorial accretion flow by determining the dominant Fourier modes of key physical quantities. In addition to the plasma density, we computed the Fourier expansion coefficients of the mass accretion rate per solid angle, $f_{M}$, and integrated the power of each mode over the duration of the flux eruption event. For the main high-cadence event analyzed in the text, our results show that the equatorial distributions of both $\rho$ and $f_{M}$ are dominated by low azimuthal number modes. Specifically, the mode with the largest time-integrated Fourier power is the $m = 2$ mode at $r_{\mathrm{H}},\, 5\, r_{\mathrm{H}}$, and $20\, r_{\mathrm{H}}$, while at $10\, r_{\mathrm{H}}$ the $m = 3$ mode dominates. These results were obtained by analyzing the structure of the equatorial accretion flow utilizing simulation data that were output with a remarkably high frequency of $1\, t_{g}$, with this time interval corresponding to the duration of the flux eruption at $32220\, t_{g}$. Repeating the same analysis for three additional flux eruptions, at $t = 14990\, t_{g}$, $t = 21890\, t_{g}$, and $t = 34190\, t_{g}$, yields the same qualitative conclusion that the equatorial inner accretion flow is governed by low-order azimuthal structure, although the specific dominant mode varies between events and radii. Overall, the additional events support the robustness of low-$m$ non-axisymmetric structure during flux eruptions, with $m = 1$ and $m = 2$ being the most recurrent dominant modes. The data corresponding to these events were output with a lower frequency of $50\, t_{g}$, which was selected after confirming the convergence of results for the event at $32220\, t_{g}$ for this lower output frequency, as shown in Fig. \ref{convergence} in Appendix \ref{appendixA}. Overall, we found that an output frequency as low as $50\, t_{g}$ is adequate for performing a quantitative analysis of an equatorial accretion flow's structure in the MAD state that accurately captures the inner disk's variable structure and morphological characteristics of large angular scale, like the one performed in this work.

The dominance of large-scale, low-$m$ structures, especially in the vicinity of the black hole ($r \leq 5\, r_{\mathrm{H}}$) suggests that the overall morphology of the equatorial accretion flow during the flux eruption is controlled by the non-axisymmetric features induced by the reconnection-driven detachment instability mechanism, rather than by small-scale turbulence. A similar prevalence of low-$m$ modes has been reported in simulations of non-relativistic accretion onto magnetized stars, where accretion streams develop through large-scale instabilities at the magnetospheric boundary.

Although interchange and Kelvin-Helmholtz instabilities are likely active in the highly magnetized inner disk, their effects are confined to smaller spatial and temporal scales. The large-scale azimuthal structure observed in our analysis reflects the reorganization of the inner accretion flow due to the detachment instability mechanism, which triggers the reconnection-induced formation and buoyant rise of vertical magnetic flux tubes during flux eruption events. This mechanism is remarkably similar to the mechanism for the production of flares during the Newtonian accretion of matter onto protostars, proposed by \cite{takasao2019}. The strong similarities between protostellar accretion and the MAD state of accretion onto black holes suggest the existence of a common regime of accretion onto objects that do not posses their own magnetospheres, but build up variable magnetospheric boundaries due to the accumulation of magnetic flux by the magnetized matter they accrete.
		
\begin{acknowledgements}
We thank the anonymous referee for their comments and suggestions. AL acknowledges financial support by the State Scholarships Foundation (IKY) scholarship program from the proceeds of the “Nic. D. Chrysovergis” bequest. AN has been supported by the Hellenic Foundation for Research and Innovation (ELIDEK) under Grant No 23698. This work was supported by computational time granted from the National Infrastructures for Research and Technology S.A. (GRNET S.A.) in the National HPC facility - ARIS - under project ID 16033. This work was also supported by the Sectoral Development Program ($\mathrm{O\Pi \Sigma}$ 5223471) of the Ministry of Education, Religious Affairs and Sports, through the National Development Program (NDP) 2021-25.
\end{acknowledgements}

		
		\bibliographystyle{aa} % style aa.bst
		\bibliography{references} % your references Yourfile.bib
        
\onecolumn
\begin{appendix}
\section{Convergence of results with output frequency}\label{appendixA}

For the analysis of the azimuthal structure of the accretion disk during flux eruption events, we chose an output frequency of $1\, t_{g}$ for the simulation results inside the time interval corresponding to the analyzed event . This choice was made in order to obtain a high temporal resolution of the equatorial accretion flow that allowed us to examine its morphological evolution as well as the formation of vertical flux bundles in great detail. However, the computational expense of such a high output cadence makes it difficult to reproduce this analysis for many events during the simulation. For this reason, we checked the convergence of the results pertinent to the azimuthal modes of the morphological features of the disk for lower frequency outputs.

\begin{figure}[H]
    \centering

    \includegraphics[width=0.246\textwidth]{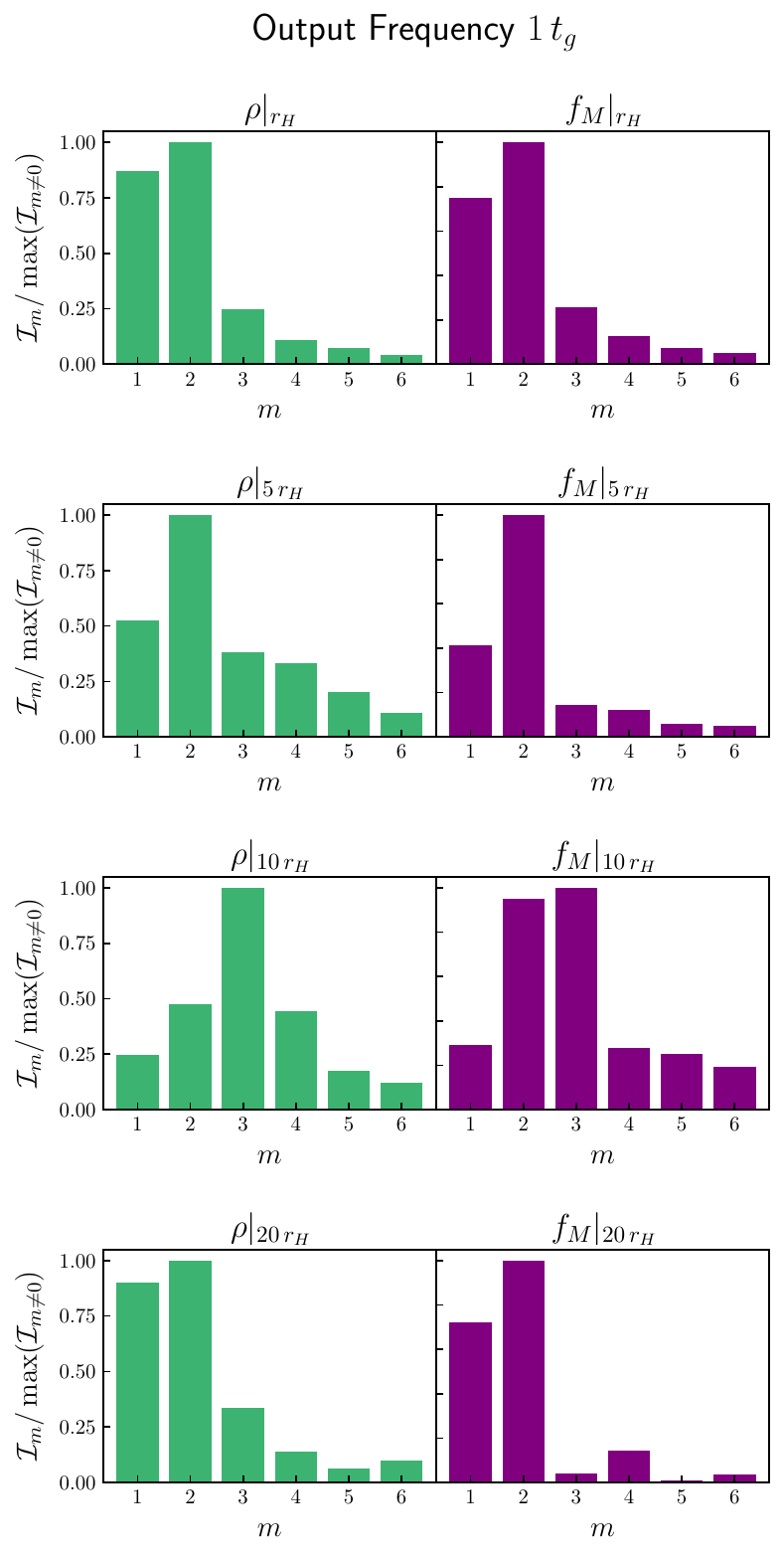}
    % \hfill
    \includegraphics[width=0.246\textwidth]{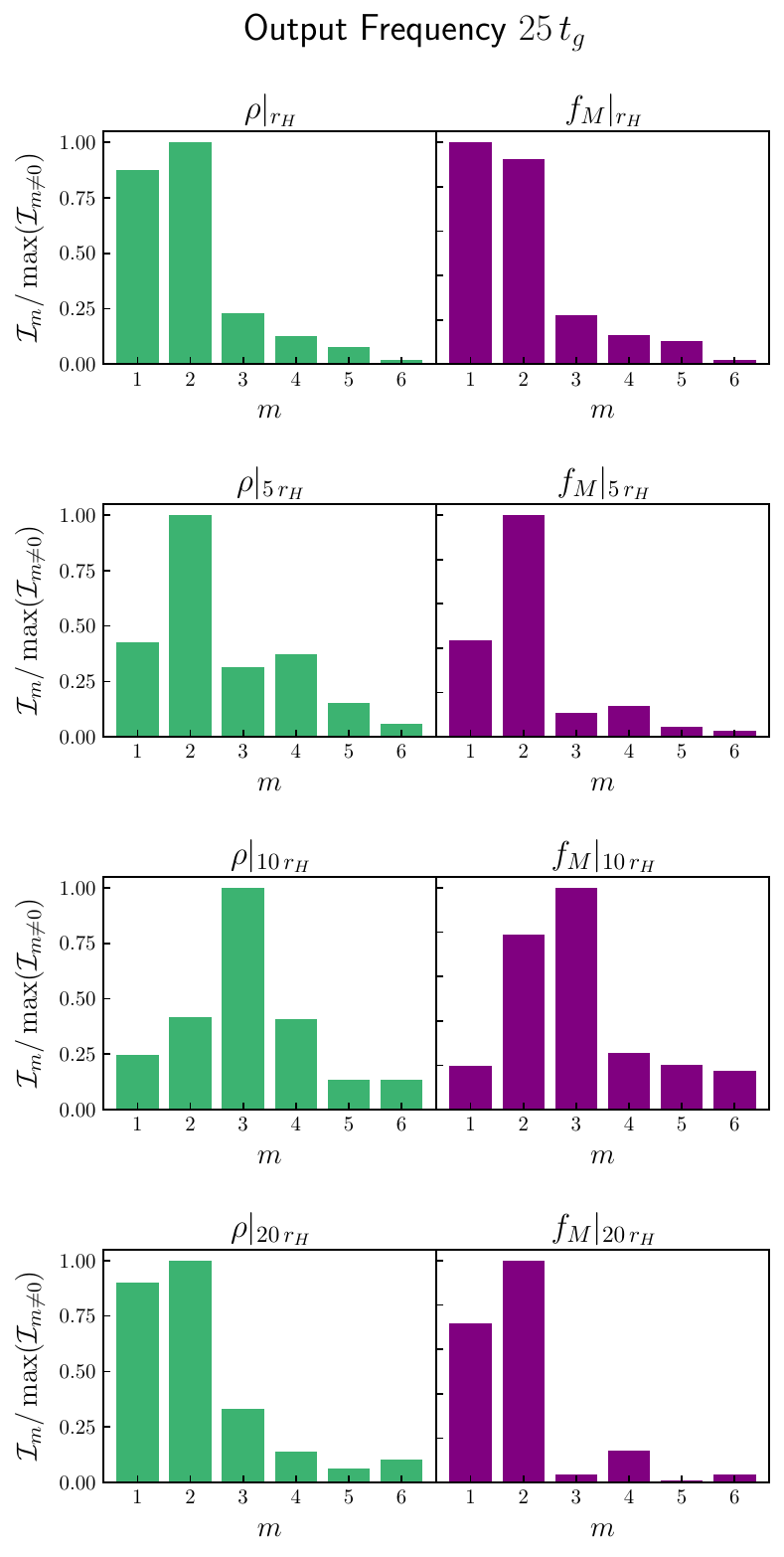}
    % \hfill
    \includegraphics[width=0.246\textwidth]{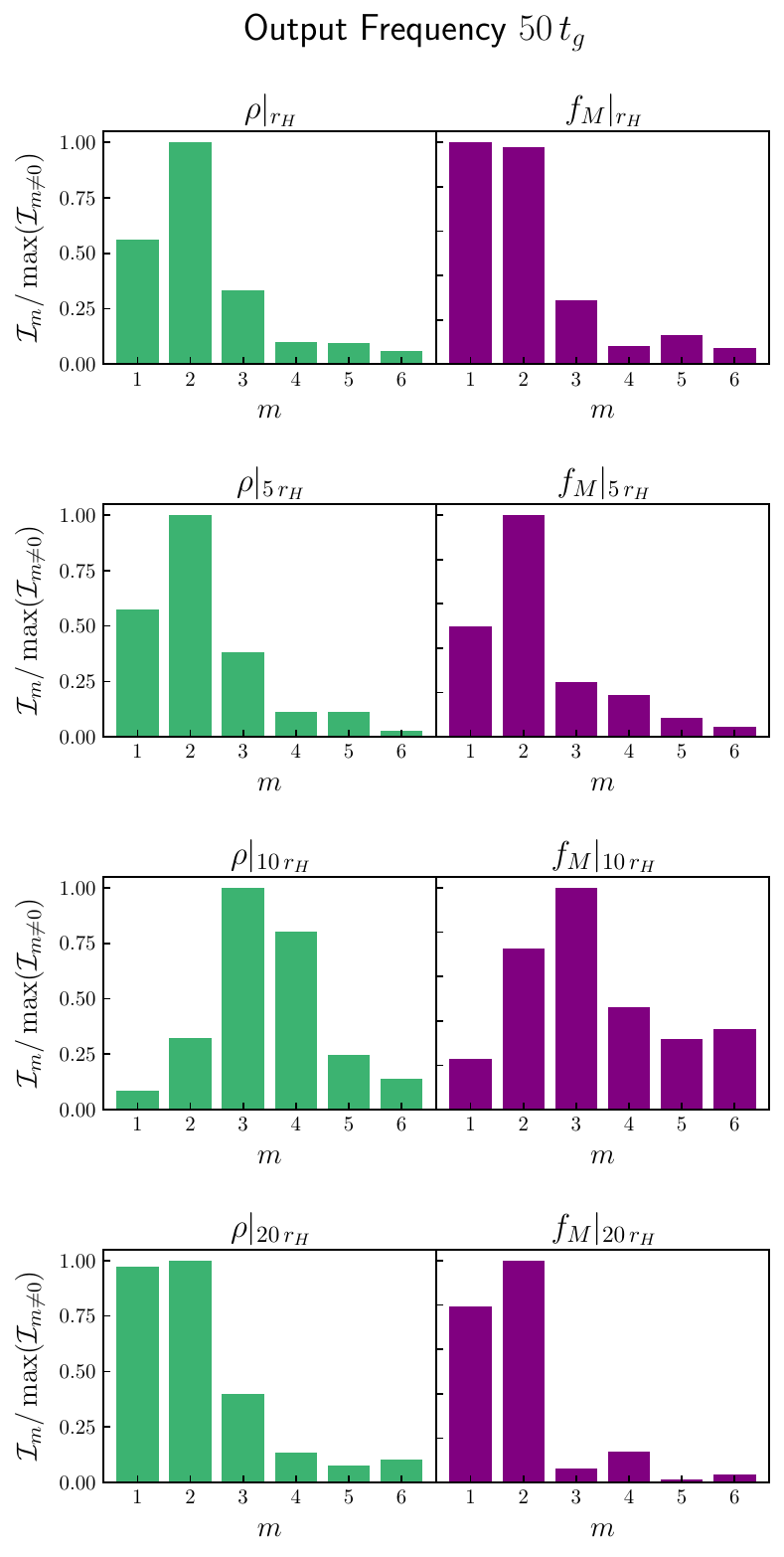}
    % \hfill
    \includegraphics[width=0.246\textwidth]{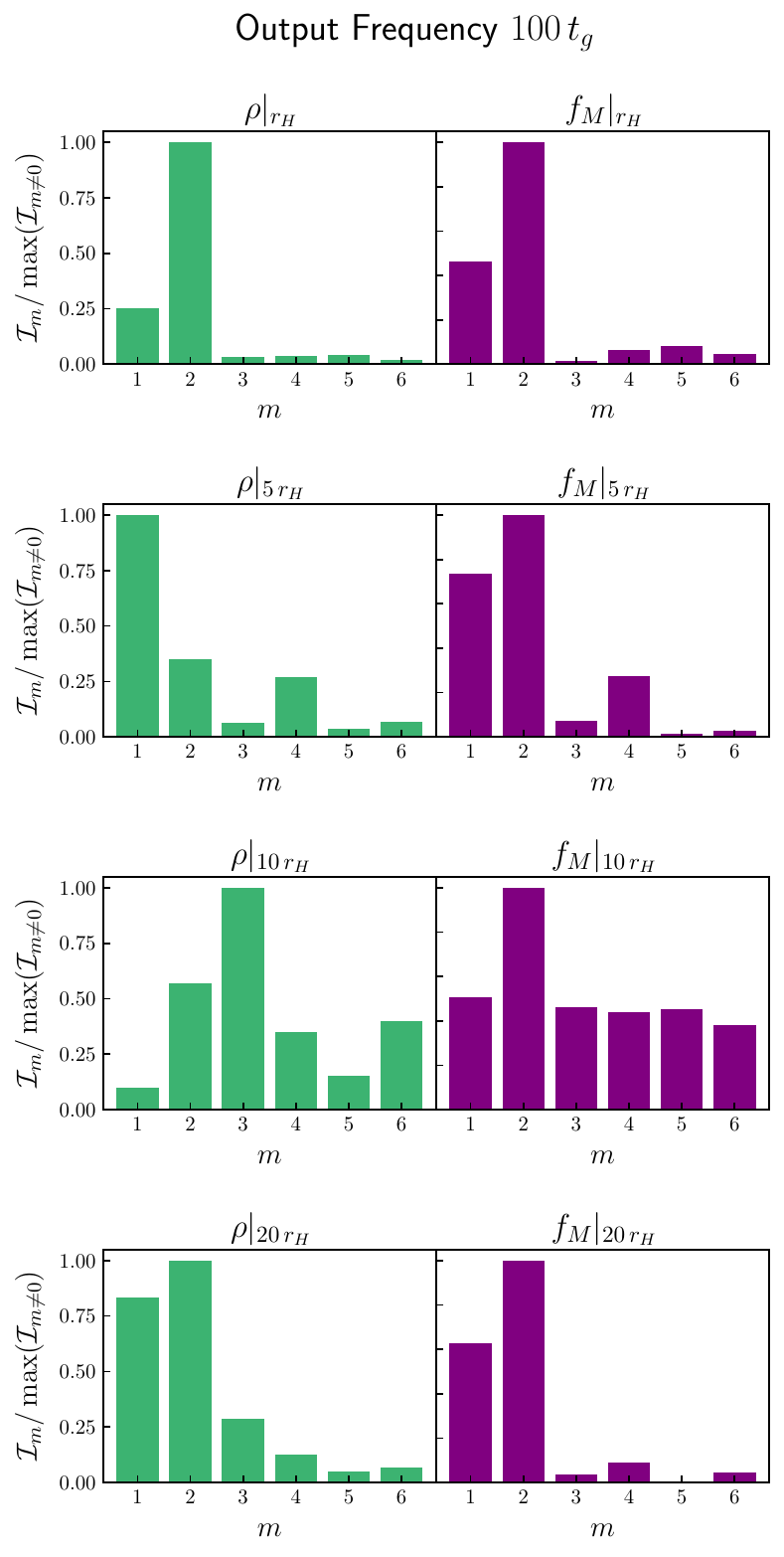}

    \caption{Time integrated squared Fourier coefficients, $\mathcal{I}_{m}$, for the equatorial plasma density, $\rho(t,\phi)$ and mass accretion rate per solid angle, $f_{M}(t,\phi)$, integrated over the entire time interval from $t = 32191\, t_{g}$ to $t = 32360\, t_{g}$. Left column: $1 \,t_{g}$ output frequency, the same as Fig.~\ref{peaks}. Middle column: $25\, t_{g}$ output frequency. Right column: $50\, t_{g}$ output cadence. Results do not vary significantly with output frequency.}
    \label{convergence}
\end{figure}

Figure \ref{convergence} displays the normalized integrals $\mathcal{I}_{m}$ of the Fourier series coefficients of $\rho$ and $f_{M}$ for an output frequency of $1\, t_{g}$ as shown in Fig. \ref{peaks}, and for lower output frequencies of $25\, t_{g}$ and $50\, t_{g}$. Results obtained by the lower output frequency data do not present significant differences to the ones obtained by the high $1\, t_{g}$ output frequency analysis, with the qualitative nature of the results remaining unchanged. We observe that in all three cases, the dominant azimuthal mode at all radii is the same, with the exception of the dominant $m$ of $f_{M}$ at $r = r_{H}$, which can be explained by the very rapid variability of the accretion flow's morphology very close to the black hole. This convergence check indicates that results regarding the azimuthal structure of the inner accretion flow obtained through data output at a significantly lower frequency are exceedingly similar to the ones obtained by our main analysis of the $1\, t_{g}$ output frequency results. Therefore, analyses built on low output frequency simulation data, as low as $50\, t_{g}$, are equally robust.

\begin{figure}[H]
    \centering
    \includegraphics[width=0.33\textwidth]{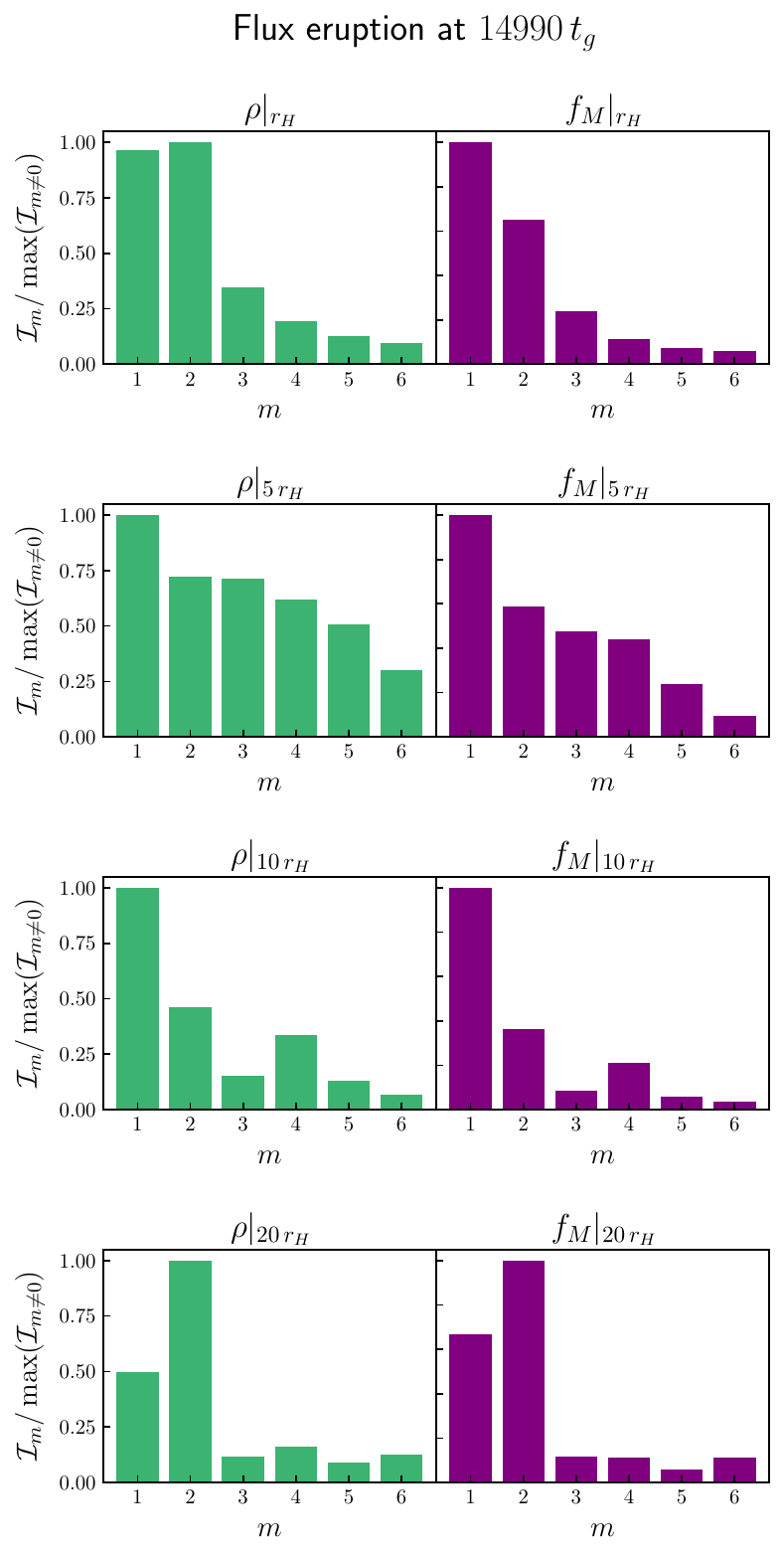}
    % \hfill
    \includegraphics[width=0.33\textwidth]{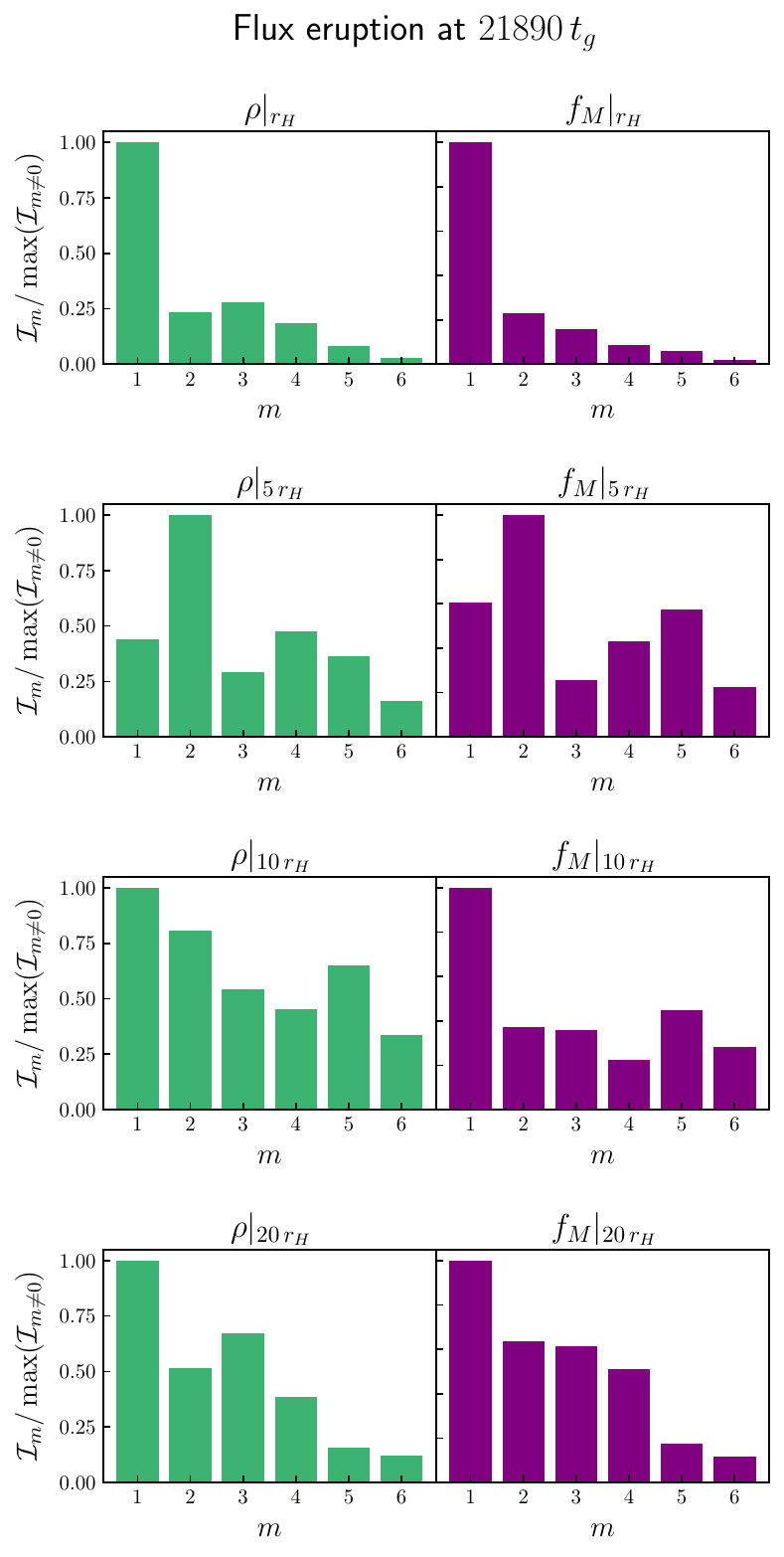}
    % \hfill
    \includegraphics[width=0.33\textwidth]{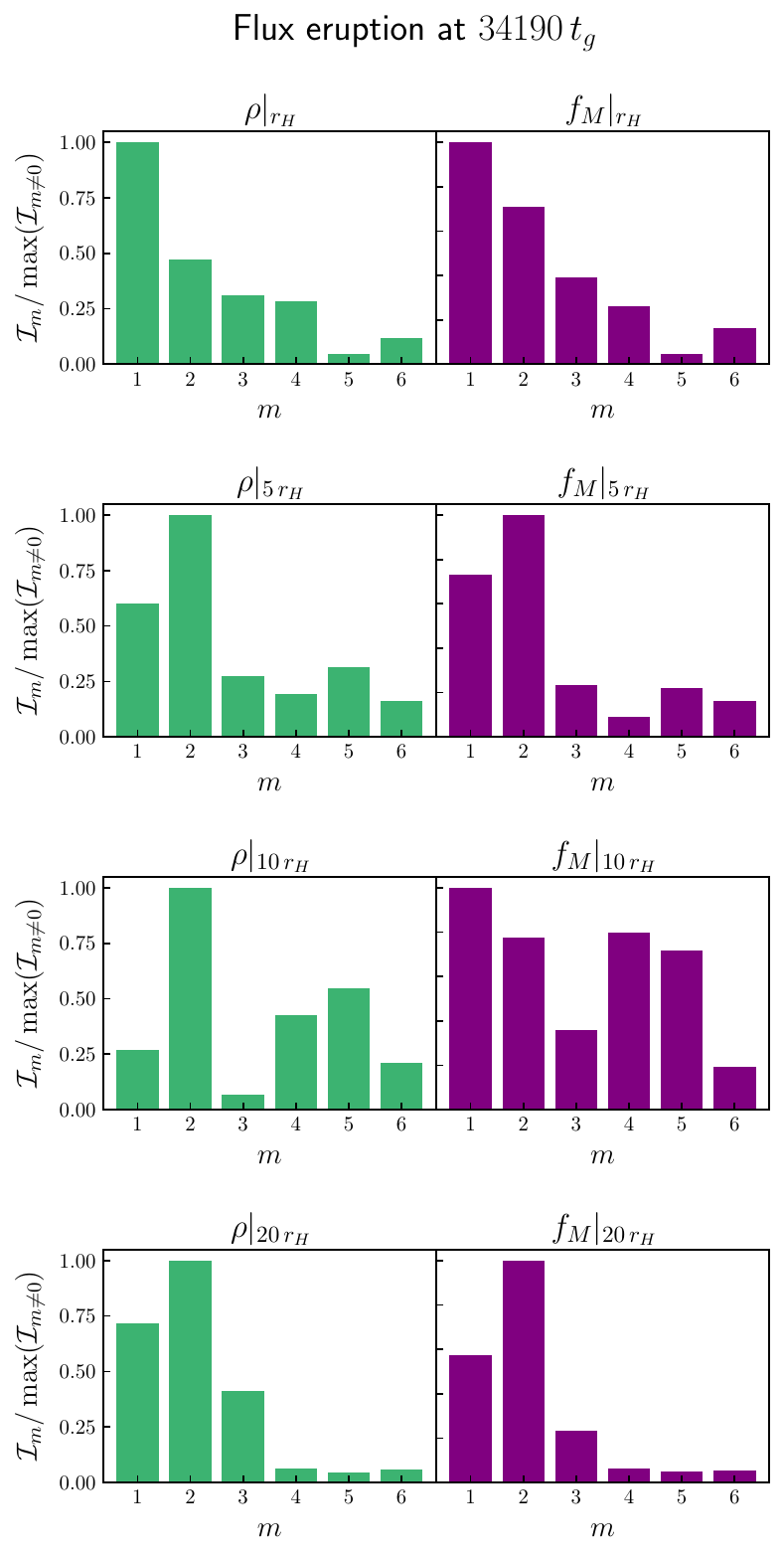}
    \caption{Time integrated squared Fourier coefficients, which express the power in each term of the expansion, $\mathcal{I}_{m}$, for the equatorial plasma density, $\rho(t,\phi)$ and mass accretion rate per solid angle, $f_{M}(t,\phi)$, for three additional flux eruption events. Left column: flux eruption event at $t = 14990\, t_{g}$. Middle column: flux eruption event at $t = 21890\, t_{g}$ Right column: flux eruption event at $34190\, t_{g}$.}
    \label{flares}
\end{figure}

We repeated our analysis of the accretion flow's structure during the flux eruption we focused on in the main text for three additional flares, at $t = 14990\, t_{g}$ and $t = 21890\, t_{g}$, before the event of the main text, and one at $t = 34190\, t_{g}$, after the event analyzed in the main text. The results are presented in Fig. \ref{flares}. The analysis of the three additional flares supports the main conclusion that the azimuthal structure of the equatorial inner accretion flow during flux eruptions is dominated by low-order modes. Although the mode with the largest time-integrated Fourier power is not identical at all radii for all events, the additional events consistently show that $m = 1$ and $m = 2$ are the most common dominant modes, while $m = 3$ becomes dominant only in a more limited subset of cases. Some of the differences between events are likely enhanced by the lower output frequency, which cannot capture the most rapid variability of the inner accretion flow equally well. Nevertheless, the qualitative agreement between the different events supports the robustness of the low-$m$ structure reported in the main text.

\end{appendix}

\end{document}